\magnification=\magstep1
\baselineskip=20 truept
\def\ni{\noindent}

\def\bn{\bigskip\noindent}
\def\mn{\medskip\noindent}
\def\sn{\smallskip\noindent}
\def\ie{{\it i.e.}}

\def\tr{\mathop{\rm tr}\nolimits}
\def\adj{\mathop{\rm adj}\nolimits}

\def\half{{\textstyle{1\over2}}}
\def\quar{{\textstyle{1\over4}}}
\def\BB{{\bf B}}
\def\CC{{\bf C}}  
\def\ii{{\rm i}}  
\def\ee{{\rm e}}  

\def\vp{\varphi}

\font\scap=cmcsc10
\font\title=cmbx10 scaled\magstep1

\def\a{\alpha}

\def\l{\lambda}

\def\s{\sigma}

\def\ddt{{{\rm d}\over {\rm d}t}}
\rightline{DTP98/59}
\vskip 1truein
\centerline{\title Lax Pairs for Integrable Lattice Systems.}

\vskip 1truein
\centerline{\scap R. S. Ward}

\bn\centerline{\it Dept of Mathematical Sciences, University of Durham,}
\centerline{\it Durham DH1 3LE, UK.}

\vskip 2truein
\ni{\bf Abstract.} This paper studies the structure of Lax pairs associated
with integrable lattice systems (where space is a one-dimensional lattice,
and time is continuous). It describes a procedure for generating examples of
such systems, and emphasizes the features that are needed to obtain
equations which are local on the spatial lattice.

\vskip 1truecm
\ni PACS 03.20.+i

\vfil\eject

\ni I. {\scap Introduction.}

\mn There has long been interest in integrable differential-difference
equations
(integrable lattice systems), especially since  the discovery of the Toda
lattice$^1$. Such systems can have direct applications, for example in
condensed-matter physics; and also occur as spatially-discrete versions
of integrable partial differential equations$^2$.  Associated with
each integrable lattice is a Lax pair, involving a matrix
$L$ which ``steps'' along the lattice, together
with another matrix $V$ which generates the time evolution. The purpose of
this paper is to investigate the structure of this Lax pair, and how it affects
the nature of the associated integrable systems.

Throughout the paper, we work with functions $\vp_n(t)$ which
depend on time $t$, and on an integer variable $n$. Such a function
will be written simply as $\vp$, its dependence on $t$ and $n$
being understood; then  $\vp_+$ denotes $\vp_{n+1}(t)$, and
$\vp_-$ denotes $\vp_{n-1}(t)$. The symbol $\Delta$ denotes the
forward-difference operator, \ie\ $\Delta\vp = \vp_+ - \vp$.
For example, the Toda lattice
equation (in first-order form) is
$$
  \ddt\vp = \psi, \qquad
  \ddt\psi = \exp\Delta\vp - \exp\Delta\vp_- \,. \eqno(1)
$$
We shall take the integer $n$ to be unrestricted
(\ie\ the lattice is infinite). Our primary interest is in systems which
are local, in the sense that the time-derivative of a variable at site
$n$ equals some expression involving the variables at sites $n-1$, $n$ and
$n+1$: \ie\ nearest neighbours only.

The point of view adopted here is that
a lattice equation is integrable if it can be written as the consistency
condition for a linear system (Lax pair) of a suitable type.
This involves two $q\times q$
matrices $L$ and $V$, the entries of which are functions of a ``spectral
parameter'' $\l$, as well as of $t$ and $n$. In what follows, we shall,
for the sake of simplicity, restrict to the
case $q=2$ (\ie\ $2\times2$ matrices).
The linear system is
$$
\eqalign{ \Psi_+ &= L\Psi\,, \cr
          \ddt\Psi &= V\Psi\,, \cr} \eqno(2)
$$
where $\Psi$ is a column 2-vector (depending on  $\l$, $t$ and $n$).
The consistency condition for (2) is
$$
   \ddt L = V_+ L - L V \,. \eqno(3)
$$
The crucial feature of (3) is that it specifies the evolution only of $L$,
and not of $V$; so in order to get a meaningful equation, $V$ has to be
determined in terms of $L$. In the next section, we shall see how this happens.
The subsequent sections illustrate how this structure can be used to generate
integrable lattice systems.  We shall see how known examples fit into this
framework; and as a new example, construct a system which couples lattice
versions of the Heisenberg ferromagnet and the derivative nonlinear
Schr\"odinger equation.

\bn II. {\scap How $L(\l)$ determines $V(\l)$.}

\mn In order to analyse the structure of $L(\l)$ and $V(\l)$,
one needs to impose some requirements on the way that they depend on
$\l$. Let us assume that $L$, $L^{-1}$ and $V$
are rational functions of $\l$, with poles at constant values of $\l$
(that is, the location of the poles does not depend on $t$ or $n$).
This is not the only possibility: for example, there is the well-known case
of the lattice Landau-Lifshitz equation$^3$, which involves elliptic functions
of $\l$. But we shall restrict to the rational case.

By making a M\"obius transformation on $\l$, we may ensure that $V(\l)$ is
finite at $\l=\infty$, \ie\ that its poles  occur at finite values of $\l$.
Furthermore, since (3) is homogeneous in $L$, we have the freedom to multiply
$L$ by a scalar function of $\l$ (not depending on $t$ or $n$). We can use this
freedom to ensure that $L$ is a (matrix)
polynomial in $\l$ which is  nonzero at each of the poles of $V$.
Let $p$ denote the degree of $L$ as a polynomial in $\l$.

Equation (3) determines the
evolution of each matrix coefficient in the polynomial $L(\l) = A\l^p
+B\l^{p-1}+\ldots+D$; so at this stage it is a set of coupled equations
for $q^2(p+1)$ functions (with $q=2$ in what follows). As was emphasized above,
the matrix $V$ has to be determined in terms of $L$,
since (3) does not specify its evolution: let us now examine how this
happens.

Assume for the time being that the poles of $V$ are all simple.
So $V$ has the form
$$
   V(\l) = \sum_{\a=1}^N V^{(\a)} (\l - \l_{\a})^{-1} + V^{(0)}\,,\eqno(4)
$$
where $V^{(0)}, V^{(1)}, \ldots, V^{(N)}$ are matrices independent of $\l$.
The general idea is that  $V^{(0)}$ is determined by a choice of gauge, whereas
each $V^{(\a)}$ for $1\leq\a\leq N$ is determined by the residue of (3) at
the pole
$\l=\l_{\a}$. Note that equations (2) and (3) are invariant under
the gauge transformations
$$ \eqalign{
\Psi &\mapsto \Lambda \Psi, \cr
L    &\mapsto  \Lambda_+ L  \Lambda^{-1}, \cr
V    &\mapsto  \Lambda V \Lambda^{-1} + 
          \Bigl(\ddt \Lambda\Bigr) \Lambda^{-1}, \cr} \eqno(5)
$$
where $\Lambda$ is a nonsingular $2\times2$ matrix depending on $t$ and $n$
(but not on $\l$).
A choice of gauge involves the following steps:
\item{(i)} choose a form for $L(\l)$ (an algebraic condition on the entries
in the matrix $L$), such that a necessary condition for this form to be
preserved under gauge transformations (5) is that $\Lambda_+ = \Lambda$;
\item{(ii)} then choose any  $V^{(0)}$ which is consistent with (i) and the
evolution equation (3).
\par\ni As an example to illustrate how this works, choose the coefficient
of $\l^p$ in  $L(\l)$ (\ie\ the leading term) to be the identity matrix.
Then the remaining gauge freedom is (5), with  $\Lambda$ independent of $n$
as required.
And  the leading term in (3) gives $ V^{(0)}_+ = V^{(0)}$, so any choice
$ V^{(0)} = V^{(0)}(t)$ then fixes the gauge completely; to obtain an
autonomous system of equations, one chooses  $V^{(0)}$ to be a constant
matrix.  This is the most straightforward choice of gauge;
for other gauges,  $V^{(0)}$ will depend on the functions
appearing in  $L(\l)$, and this dependence is in general nonlocal as we shall
see below.

Consider, next, the pole at $\l=\l_{\a}$. Clearly the
residue of the right-hand side of (3) at this pole must vanish, \ie\
$$
  V^{(\a)}_+ L^{(\a)} - L^{(\a)} V^{(\a)}=0, \eqno(6)
$$
where  $L^{(\a)} = L(\l_{\a})$.
The idea is that this constraint determines  $V^{(\a)}$. Since  $L^{(\a)}$
is a non-zero $2\times2$ matrix, there are two gauge-invariant possibilities:
the rank of  $L^{(\a)}$ could be either 2 or 1. The next section deals with the
rank 2 case; thereafter, we concentrate on the rank 1 case.

\bn III. {\scap Rank 2 Case: Nonlocal Systems.}

\mn If  $L^{(\a)}$ is invertible, then (6) is a difference equation
$$
 V^{(\a)}_+  = L^{(\a)} V^{(\a)} L^{(\a)-1},
$$
which determines  $V^{(\a)}_n$  in terms of (say) $L^{(\a)}_{n-1}$,
$L^{(\a)}_{n-2},\ldots$ and $V^{(\a)}_{-\infty}$.
In other words, $V^{(\a)}$ is a nonlocal function of the entries in
$L^{(\a)}$.

To obtain a simple example which illustrates this case,
take $p=N=1$. Without loss of generality, we may set $\l_1=0$. Write
$L = A\l + B$, where $B$ is invertible; and choose a gauge by specifying
$$
  A = \pmatrix{0 &1 \cr 1 &0 \cr }
$$
and $V^{(0)}  = 0$. So $V(\l) =  V^{(1)}/\l$, where
$V^{(1)}$ is the solution of the difference equation
$$
  V^{(1)}_+ = B V^{(1)} B^{-1}. \eqno(7)
$$
In general, the $2\times2$ matrix $B$ contains four functions; let us effect
a reduction to one function $y_n(t)$ by taking $B$ to have the form
$$
  B = \pmatrix{\ee^y &0 \cr 0 &\ee^{-y} \cr}. \eqno(8)
$$
In order for the reduction to be consistent, we need  $V^{(1)}_{-\infty}$
to have the form
$$
   V^{(1)}_{-\infty} = \pmatrix{0 &b \cr c &0 \cr}
$$
(plus a multiple of the identity, which has no eventual effect).
The resulting system of equations for $y$ is
$$
  \ddt y = b \exp(Y) - c \exp(-Y), \eqno(9)
$$
where $Y_n(t)$ is given by the non-local expression
$$
  Y_n = y_n + 2 \sum_{k=-\infty}^{n-1} y_k .
$$
This equation can be transformed to a form which looks local, by writing
$y_n = \phi_n - \phi_{n+1}$. In terms of $\phi_n(t)$, eqn (9) becomes
$$
 \ddt(\phi_+ - \phi) = c\exp(\phi_+ + \phi) -  b\exp(-\phi_+ - \phi).\eqno(10)
$$
If (for example) $b=0$, then (10) is a differential-difference version of
the Liouville equation
$$
\phi_{tx} = \exp(2\phi). \eqno(11)
$$
To get (11) from (10), we interpret $c$ as the lattice spacing, put $x=nc$,
and take the continuum limit $c\to0$. Similarly, putting $b=c$ in (10)
gives  a differential-difference version of the sine-Gordon equation$^4$.
But for these lattice Liouville and sine-Gordon systems,
$x$ and $t$ (or $n$ and $t$) do not represent space and time;
rather they are characteristic (null) coordinates. In particular, one cannot
specify arbitrary initial data at $t=0$. In these lattice equations, one of
the characteristic coordinates has become discrete, while the other (namely
$t$) remains continuous.

This example can be generalized in several directions, as follows.
If one does not make the reduction (8), then one obtains an equation
for the matrix $B$. Choosing the slightly different gauge $A = {\bf I}$
(the identity matrix) and  $V^{(0)}  = 0$, leads to
$$
  \ddt B - \Delta V^{(1)} = 0 \eqno(12)
$$
together with (7). Now (7) and (12) ensure that there exists a matrix
$R_n(t)$ with
$$
  B = 1 + R_+ R^{-1}, \qquad V^{(1)} = -\bigl(\ddt R\bigr) R^{-1}.
$$
Then (12) becomes
$$
  \ddt( R_+ R^{-1}) + \Delta\bigl[ \bigl(\ddt R\bigr) R^{-1} \bigr] = 0.
                        \eqno(13)
$$
This is a differential-difference version of the principal chiral equation
$$
   (R_x  R^{-1})_t +  (R_t  R^{-1})_x = 0 \eqno(14)
$$
in which, as before, one of the characteristic coordinates has become discrete.

This chiral equation generalizes, of course, to larger matrices
($q>2$). Similarly, the Liouville and sine-Gordon examples generalize to
differential-difference versions of other Toda field equations.

Finally, it might be noted that
there are  difference-difference versions of the principal chiral$^5$
and Toda field$^6$ equations in which both characteristic coordinates
(here $x$ and $t$) become discrete. Another, very general, example of this
type is the Hirota bilinear difference equation$^7$.

So systems with $\det L^{(\a)} \neq 0$ may be thought of as time-evolution
equations which are nonlocal on the spatial lattice, or as equations where
a characteristic coordinate (neither space nor time) has become discrete.
To get local evolution equations, it is necessary for each $L^{(\a)}$ to have
rank 1. From now on, we shall concentrate on this case.


\bn IV. {\scap Rank 1 Case: Local Systems.}

\mn Given that  $L^{(\a)}$ has rank 1, the constraint  (6) may be solved as
follows. Write $K^{(\a)} = L_-^{(\a)}L^{(\a)}$. Assuming that $\tr K^{(\a)}$ is
non-zero, the general solution of (6) is
$$
   V^{(\a)} = {1\over\tr K^{(\a)}}\bigl[f^{(\a)} K^{(\a)} +
                  g^{(\a)}\adj K^{(\a)} \bigr], \eqno(15)
$$
where $\adj K^{(\a)}$ denotes the adjoint matrix of $K^{(\a)}$, and where
$f^{(\a)}$
and $g^{(\a)}$ are scalar functions, with $f_+^{(\a)} = f^{(\a)}$.
So the constraint (6) does not determine
$V^{(1)},\ldots,V^{(N)}$ uniquely: in particular one has the arbitrary
functions  $g^{(\a)}$. However, there is a further constraint, namely that
the condition $\det L^{(\a)} = 0$ has to be preserved by the evolution (3).
This gives equations on the  $g^{(\a)}$, which are precisely that they are
constant on the lattice, just as the $f^{(\a)}$ are:  $g_+^{(\a)} = g^{(\a)}$.
Then (15) can be re-written as 
$$
 V^{(\a)} = { c^{(\a)} \over\tr K^{(\a)}}K^{(\a)}
                     +  d^{(\a)} {\bf I},
$$
where $c^{(\a)}$ and  $d^{(\a)}$ are functions of $t$ only.
It is clear that the  $d^{(\a)}$ term will not contribute in
the evolution equations, and so only the $c^{(\a)}$ remain;
we may as well set  $d^{(\a)} = 0$, and take
$$
 V^{(\a)} = { c^{(\a)} \over\tr K^{(\a)}}K^{(\a)}. \eqno(16)
$$
At this stage, the  $c^{(\a)}$ could still be functions of
time $t$; for simplicity, let us take  them to be constants.
One point to note about (16) is that $V^{(\a)}$ is
local:  $V^{(\a)}_n$ is expressed in terms of $L_n$ and $L_{n-1}$.

So to obtain local evolution equations with  $2\times2$ matrix Lax pairs,
one first specifies the integer $p$ (the degree of $L(\l)$);
the integer $N$ appearing in (4) equals $2p$, since
the $\l_{\a}$ all have to be roots of $\det L(\l)$. The matrices $V^{(\a)}$
for $1\leq\a\leq2p$ are given by (16), and involve the $2p$
constants $c^{(\a)}$.  Finally, there is the choice of gauge, which determines
$V^{(0)}$. In general,  $V^{(0)}$ turns
out to be nonlocal, and special gauge choices are needed to ensure that it
is local.

One can relate all this to the $r$-matrix description (see, for example,
reference 3; and also reference 8 which addresses the construction of an
$r$-matrix from a given Lax pair). Suppose one has an $L(\l)$, a Poisson
bracket and an $r$-matrix such that the Fundamental Poisson Bracket Relations
are satisfied. Suppose also that there exist $\l_1,\ldots,\l_N$ such that
$\det L(\l_{\a}) = 0$ for each $\a$. Let $\tau(\l)$ be the trace of the
monodromy
matrix $\ldots L_2(\l) L_1(\l) L_0(\l) L_{-1}(\l)\ldots$ (which propagates
from $n=-\infty$ to  $n=+\infty$). Then
$$
  H = \sum_{\a=1}^N c^{(\a)} \log \tau(\l_{\a})
$$
is a local Hamiltonian, and the corresponding Hamiltonian equations are just
(3); the constants  $c^{(\a)}$ are the same as those appearing in (16).
The problem from this point of view is to choose  $L(\l)$, in a suitable
gauge, such that a compatible $r$-matrix structure exists.

More generally, one wants  $\tau(\l)$ to be conserved in time, for all $\l$
--- this then gives infinitely many conserved quantities.  If one has a
Lax pair (2) and boundary conditions which imply that
$V_{+\infty} = V_{-\infty}$, then  $\tau(\l)$ is indeed conserved.  When
$V(\l)$ dependes locally on the fields, then then it is easy to ensure
that this condition is met; if, on the other hand, $V$ is nonlocal, then
conservation of  $\tau(\l)$ is not guaranteed.  This is one reason why
locality is desirable, in the present context.

If $p = 1$ (in other words, $L(\l)$ is linear in $\l$), then
$\det L(\l)$ is a quadratic polynomial in $\l$, the roots of which are
$\l_1$ and $\l_2$. For the time being, let us assume that these roots are
distinct; and by translating $\l$ set $\l_1 = 1$ and $\l_2 = -1$. It follows
that  $L(\l)$ has the form
$$
  L(\l) = \half(\l + 1) L^{(1)} -\half(\l - 1) L^{(2)}, \eqno(17)
$$
where  $L^{(1)}$ and $L^{(2)}$ are $2\times2$ matrices each having zero
determinant. So the entries in  $L^{(\a)}$ involve six independent functions
of $t$ and $n$ (in effect, the requirement that $\l_1$ and $\l_2$ be constant
has reduced the number of functions in $L(\l)$ from eight to six). The two 
$L^{(\a)}$
satisfy evolution equations obtained by expanding (3): these are
$$
\ddt  L^{(\a)} = V_+^{(0)} L^{(\a)} -  L^{(\a)} V^{(0)} - \half\Xi\,, \eqno(18)
$$
where $\Xi = V_+^{(1)} L^{(2)} -  L^{(2)} V^{(1)} -
V_+^{(2)} L^{(1)} +  L^{(1)} V^{(2)}$.  The gauge choice
reduces the number of functions by four (since $\Lambda$ contains four
entries), and we end up with a system involving two functions.
A number of examples of this type are described in the following section.


\bn V. {\scap Some $p=1$ Examples.}

\mn This section exhibits some systems of the type described
in the previous section.  Such examples are simple to generate; but before
doing so, we should ask when two lattice equations are to be regarded as being
``the same''.  More specifically, is there an appropriate equivalence relation
on the set of all such systems?  Certainly such an equivalence would include
gauge transformations in which $\Lambda$ was constant; and strictly-local
redefinitions of the functions appearing in $L$ (\ie\ the new functions
at lattice-site $n$ depend on the old functions at site $n$ only).
However, it is customary to allow more general transformations than just
these.   A well-known case is that of the Toda lattice (1).   If one replaces
$\vp$ by
$$
   r = \Delta\vp_- = \vp - \vp_- \,, \eqno(19)
$$
then (1) becomes
$$
  {{\rm d}^2 \over {\rm d}t^2} r = \ee^{r_+} - 2\ee^r + \ee^{r_-}\,; \eqno(20)
$$
and this is regarded as simply another form of the Toda lattice equation.

But any equivalence relation which admitted (1) and (20) to the same class,
would also have to allow the transformations $\vp\mapsto\Delta^k \vp$
for all integers $k$ (negative as well as positive).  If one allows such
highly-nonlocal transformations, then ones ends up with rather few
equivalence classes;
in fact, one might as well transform to action-angle variables,
and say that the ``only'' integrable lattice is linear.  Clearly this is
inappropriate.

The point of this argument is to conclude that there is no useful equivalence
relation on integrable systems of the type that we are considering (unless
we insist that (1) and (20) are to be regarded as distinct).  This means that
the task of listing such systems in a systematic way is not really
well-defined.  The best that one can do is to exhibit examples, and indicate
how they are related to one another.


\mn{\it Example (i).} Choose
a gauge such that  $L^{(1)} - L^{(2)} = 2{\bf I}$. This is the gauge
which was mentioned as an example in section II. As was
remarked there, the gauge is then fixed completely by specifying some
$V^{(0)}(t)$. The simplest choice is to set  $V^{(0)} = 0$. 
Note that  $L^{(1)}$ and $L^{(2)}$ must have the form
$$
L^{(1)} = {\bf I} + M, \qquad L^{(2)} = -{\bf I} + M,
$$
where $M$ is trace-free and $\det M = -1$. So we may write
$M = {\bf f\cdot\s}$, where $\s_1$,  $\s_2$ and  $\s_3$ are the Pauli
matrices, and  ${\bf f = f}_n(t)$ is a unit 3-vector. The dot denotes
the usual 3-dimensional Euclidean scalar product (and  $\wedge$ below will
denote the vector product). The evolution equation for ${\bf f}$, derived
from (18), is then
$$
\ddt{\bf f} = \Delta\bigl[(1 + {\bf f_-\cdot f})^{-1}
            (\mu{\bf f_-} + \mu{\bf f} + \nu{\bf f_-\wedge f})\bigr],
            \eqno(21)
$$
where $\mu = \half(c^{(1)} - c^{(2)})$ and
$\nu = \half\ii(c^{(1)} + c^{(2)})$ are constants.
Equation (21), then, is an integrable equation for the unit-vector function
$ {\bf f}={\bf f}_n(t)$. If the parameter $\nu$ is non-zero, it can be set to
unity by scaling $t$; so the system effectively depends on the single
parameter $\mu$. The case  $\mu=0$ is the ``Lattice Heisenberg Model''$^{3,9}$,
so-called because it has the equation of the Heisenberg
ferromagnet as a continuum limit. Indeed, if we set $\nu=2/h^2$,
$\mu=\hat\mu/h$, and let $h\to0$, then (21) becomes
$$
  {\bf f}_t = \hat\mu{\bf f}_x + {\bf f}\wedge{\bf f}_{xx}; \eqno(22)
$$
the Heisenberg model corresponds to $\hat\mu=0$. A slightly different choice
of gauge, namely one in which $V^{(0)}$ is a nonzero constant matrix, yields a
lattice nonlinear Schr\"odinger equation (different from the one in the next
example). This is the lattice counterpart
of the well-known gauge equivalence of the  nonlinear Schr\"odinger
and Heisenberg systems.

\mn{\it Example (ii).} Choose a gauge such that
$$
  L^{(1)} = \pmatrix{1 &0 \cr u &0 \cr}\,, \qquad
  L^{(2)} = -\pmatrix{0 &v \cr 0 &1 \cr}\,, \eqno(23)
$$
where $u$ and $v$ are functions of $t$ and $n$. The remaining gauge freedom
has $\Lambda$ being a diagonal matrix function of $t$ only.
Substituting (23) into the evolution equation (18) determines
$V^{(0)}$, as
$$
V^{(0)} = \half\pmatrix{-c^{(1)}vu_- & c^{(1)}v + c^{(2)}v_- \cr
                     c^{(2)}u + c^{(1)}u_- & -c^{(2)}uv_- \cr}
$$
plus a diagonal matrix function of $t$, which by the residual gauge freedom may
be set to zero. In addition, (18) gives the equations for $u$ and
$v$, namely
$$\eqalign{
 \ddt u &= \half( c^{(2)} \Delta u +  c^{(1)} \Delta u_- +  c^{(1)} vuu_-
                                       -  c^{(2)} vuu_+), \cr
 \ddt v &= \half( c^{(1)} \Delta v +  c^{(2)} \Delta v_- +  c^{(2)} uvv_-
                                       -  c^{(1)} uvv_+). \cr}\eqno(24)
$$
This is exactly the Ablowitz-Ladik system$^2$; their $L$-operator is
slightly different from the one presented here (it is, in effect, quadratic
rather than linear in $\l$); but it is easily seen to be equivalent. In
particular, if we choose  $c^{(1)} = 2\ii =  -c^{(2)}$, and impose the
(consistent) reduction $v = \pm u^*$, then (24) reduces to a lattice
nonlinear Schr\"odinger equation
$$
\ii \ddt u = u_+ - 2u + u_- \mp u u^* ( u_+ +  u_-)\,.
$$

\mn{\it Example (iii).} Here we choose a gauge such that $L^{(1)}$ is constant.
Without loss of generality, we may take  $L^{(1)}$ to be
$$ 
 L^{(1)} = \pmatrix{1 &0 \cr 0 &0 \cr}\,, \eqno(25)
$$
The most general form for the matrix $L^{(2)}$ is
$$ 
 L^{(2)} = \pmatrix{uv &uw \cr v &w \cr}\,, \eqno(26)
$$
where $u$, $v$ and $w$ are functions of $n$ and $t$. The evolution equation
(18) for $L^{(1)}$ implies that $V^{(0)}$ must have the form
$$
V^{(0)} = -{c^{(1)}\over2} \pmatrix{0 & uw \cr v_- &0 \cr}
         - { c^{(2)}\over2(v u_- + w)}
            \pmatrix{v u_-  &w u_- \cr  v  &0 \cr}
         + \pmatrix{A & 0 \cr 0 & D \cr}\,, \eqno(27)
$$
where $\Delta A = 0$ (\ie\ $A$ depends only on $t$).  The residual gauge
freedom, \ie\ that which preserves (25), is (5) with
$$
  \Lambda = \pmatrix{ f(t) & 0 \cr 0 & g(n,t) \cr } \,;
$$
this has to be used to determine $A$ and $D$, and to eliminate one of the three
functions $u$, $v$ or $w$. In fact, the role of $f(t)$ is simply to fix $A$:
let us choose $A = \half c^{(1)}$. The remaining freedom now is
$$\eqalign{
  u &\mapsto  g_+^{-1} u, \cr
  v &\mapsto  g_+ v, \cr
  w &\mapsto  g^{-1}g_+ w. \cr}\eqno(28)
$$

Equation (18) for  $L^{(2)}$ gives equations for  $u$, $v$ and $w$, one form
of which is
$$\eqalign{
 \ddt \log v &= D_+ + \half c^{(1)} (w v_-/v - uv)
                + \half c^{(2)}{v_+ + v w_+ \over v(u v_+ + w_+)}, \cr
 \ddt \log w &= \Delta\biggl(D 
             - \half c^{(2)}{v u_- \over v u_- + w}\biggr), \cr
 \ddt (uv) &= \Delta\biggl(- \half c^{(1)}uwv_-
             + \half c^{(2)}{v u_- \over v u_- + w}\biggr), \cr
 \ddt \log(uw) &= -D - \half c^{(1)}(u_+ w_+ /u - uv)
         - \half c^{(2)}{uw + u_- \over u(v u_- + w)}. \cr
                                                       }\eqno(29)
$$
(Any three of these equations implies the fourth.)  We see from (28) that
in order to remove the remaining gauge freedom, \ie\ fix $g$ (at least up to
a function of $t$), there are three possibilities.  Namely, we can specify
either
$w$, or $uw$, or $v$ as a function of the gauge-invariant combination $uv$.
This in turn will determine $D$, and hence $V^{(0)}$, on eliminating the
relevant variable from (29). For example, specifying $uw$ as a function of
$uv$ will give a local formula for $D$. Similarly, $v$ can be specified as
any function of $uv$.  But if we impose $w = F(uv)$, then we need
$F$ to be an exponential in order to get a local expression for $D$.
This illustrates the way in which some choices of gauge lead to a nonlocal
expression for $V^{(0)}$.

As an example, let us take the
gauge $w = {\rm constant}$. Choose $w = -1$, and write $u = -\ee^x$,
$uv = y$. Then (29) reduces to the system
$$\eqalign{
  \ddt y &= \half c^{(1)}\Delta(y_-\ee^{\Delta x_-}) + 
             \half c^{(2)}\Delta\biggl({y \over y - \exp\Delta x_-}\biggr), \cr
  \ddt x &= \half c^{(1)} (y + \ee^{\Delta x}) -
           {c^{(2)} \over 2(y - \exp\Delta x_-)}. \cr
} \eqno(30)
$$
This is a version of the relativistic Toda lattice$^{10--14}$.
\mn{\it Example (iv).} The gauge choice
$$
L^{(1)} = \pmatrix{-1 &\ee^x \cr 0 &0 \cr}, \qquad
L^{(2)} = \pmatrix{\ee^y &0 \cr -k\ee^{y-x} &0 \cr},
$$
where $k$ is a constant, gives
$$
V^{(0)} = {c^{(1)}\over2} \pmatrix{1-k\ee^{y_- + \Delta x_-} & -\ee^x \cr
                                   -k\ee^{y_--x_-} &0 \cr}
         + {c^{(2)}\over2}\pmatrix{0 &\ee^{x-y} \cr
                              k\ee^{-x_-} & 1 - k\ee^{-y+\Delta x_-} \cr}\,,
$$
and again leads to the relativistic Toda system$^{10,14,15}$, this time in
the form
$$\eqalign{
  \ddt y &= -\half c^{(1)}k\Delta(\ee^{y_-+\Delta x_-}) -
             \half c^{(2)}k\Delta(\ee^{-y+ \Delta x_-}), \cr
  \ddt x &= -\half c^{(1)}\ee^y (1 + k\ee^{\Delta x}) +
           \half c^{(2)}\ee^{-y}(1 + k\ee^{\Delta x_-}). \cr
} \eqno(31)
$$
The limit $k\to0$ gives the Toda system (1); it is worth examining this in
more detail. In order to get (1), one may replace the variables $x$ and $y$
by $\vp$ and $\psi$, where
$$
\ee^x = -\sqrt{k} \ee^{\vp}, \qquad \ee^y = -1 + \sqrt{k}\psi,
$$
and set  $c^{(1)} = c^{(2)} = -1/\sqrt{k}$. Then the  $k\to0$ limit of (31) is 
indeed (1). But since $c^{(\a)}\to\infty$ in this limit, we need to
re-interpret the associated Lax pair. The way to get a well-behaved limit 
is to replace $\l$ by $2\l/\sqrt{k}$. The roots of $\det L(\l)$ now occur at
$\l = \pm\half\sqrt{k}$, and so in the $k\to0$ limit they coincide. In fact,
when $k=0$ we have
$$
  L(\l) = -\pmatrix{1 + \psi\l & \ee^{\vp}\l \cr
                    -\ee^{-\vp} \l &0 \cr},
$$
and $\det L(\l)$ has a double zero at $\l = 0$ (cf.\ ref 3). The corresponding
expression for $V(\l)$ is obtained by taking the  $k\to0$ limit after first
subtracting a constant multiple of the identity matrix: this yields
$$
  V(\l) = \pmatrix{0 & -\ee^{\vp}\cr \ee^{-\vp_-} &\l^{-1}\cr}.
$$

\bn VI. {\scap Some $p=2$ Examples.}

\mn  In the $p=2$ case, $L(\l)$ has the form
$L(\l) = A\l^2 + B\l + C$, where $A$, $B$ and $C$ are $2\times2$ matrices;
so to begin with, one has twelve functions of $n$ and $t$.  The requirement
that the zeros $\l_\a$ of the quartic polynomial  $\det L(\l)$ be constant
imposes four equations on these functions, and choice of gauge imposes a
further four, so one is left with four independent functions.  In other words,
the generic system in this $p=2$, $q=2$ case is a system of coupled
evolution equations for four lattice variables.

Reductions of such systems, so that fewer functions are involved, are of
course possible.  One example that has been known for some time is
a lattice version of the sine-Gordon equation
in which space is discrete and time continuous (by contrast with the version
mentioned in section III). Here the $L$-operator has the form$^{16,3}$
$$
 L = \pmatrix{\l f(\vp) \ee^{\ii\eta} 
                      & \quar h(\ee^{-\ii\vp/2}-\l^2\ee^{\ii\vp/2}) \cr
               \quar h(\l^2\ee^{-\ii\vp/2}-\ee^{\ii\vp/2})
                      & \l f(\vp) \ee^{-\ii\eta} \cr},
$$
where $\vp$ and $\eta$ are functions of $n$ and $t$, $h$ is a constant
corresponding to the lattice spacing, and
$f(\vp) = (1+{1\over8}h^2\cos\vp)^{1/2}$.
The resulting integrable lattice has sine-Gordon in ``laboratory coordinates''
as a continuum limit: if we replace $n$ by $x=nh$ and let $h\to0$, then $\vp$
satisfies $\vp_{tt} - \vp_{xx} + \sin\vp = 0$.

In order to obtain another example, let us choose a different gauge, namely
$A = {\bf I}$, $V^{(0)} = 0$.  On $B$ and $C$ we impose the four constraints
$\tr B = 0$, $\det C = l$ constant, $\tr(BC) = 0$, and $\tr C + \det B = -2k$
constant.  It then follows that $\det L(\l) = \l^4 - 2k\l^2 + l$ has
constant zeros.  The matrices $B$ and $C$ now involve four independent
functions, and their evolution is given by
$$\eqalign{
 \ddt B &= \Delta Q, \cr
 \ddt C &= R_+ C - CR, \cr
}\eqno(32)
$$
where $Q = \sum_\a V^{(\a)}$,  $R = -\sum_\a \l_\a^{-1} V^{(\a)}$, and the
$V^{(\a)}$ are constructed as in (16).  There are four parameters, namely
the $c^{(\a)}$.

We can get an idea of what this system represents by looking at a continuum
limit.  To keep things simple, we assign particular values to the parameters,
and this leads to the following continuum integrable system.

Let \BB\ and \CC\ be 3-vectors, functions of $x$ and $t$, satisfying the
constraints $\BB\cdot\CC = 0$ and $\CC\cdot\CC = 1$.  Their time-evolution
is given by
$$\eqalign{
\BB_t &= (\CC\wedge\BB_x)_x - \{(\CC\wedge\CC_x\cdot\BB)\CC\}_x
          + \half\{(\BB\cdot\BB)\BB\}_x \,, \cr
\CC_t &= (\CC\wedge\CC_x)_x + (\CC\wedge\CC_x\cdot\BB)\BB\wedge\CC
          + \half(\BB\cdot\BB)\CC_x \,. \cr
}\eqno(33)
$$
This is integrable: it has a Lax pair of the form $\Psi_x = U\Psi$,
$\Psi_t = V\Psi$, where
$U = -\half\ii(\BB\l^{-1} + \CC\l^{-2})\cdot{\bf\sigma}$ corresponds to
$L(\l)$, and $V = \sum_{k=1}^4 V_k \l^{-k}$ is a limiting version of the
$V(\l)$ of the lattice system.

The equations (33) have two obvious reductions: if $\BB=0$, then we are left
with the Heisenberg ferromagnet equation for \CC; while if \CC\ is a constant
unit vector, then \BB\ satisfies the derivative nonlinear Schr\"odinger
equation.  So (33) may be viewed as a coupled Heisenberg-DNLS system; and
(32) is a spatially-discrete version of this coupled system.

\bn VII. {\scap Concluding Remarks.}

\mn It is clear that the examples given above
provide only a very small sample of integrable lattice systems. One may
envisage a classification involving the three integers $q$ (the size of the
matrices $L$ and $V$), $p$ (the degree of $L(\l)$) and $r$ (the maximum
order of the poles of $V(\l)$).  But in view of the remarks at the beginning
of section V, a complete classification would require a way of dealing with
the problem of equivalence.

We conclude with a remark on higher values of $r$.
One can take a given $L(\l)$ (thereby fixing $q$ and $p$), and allow
$r>1$: in other words, higher-order poles in  $V(\l)$. This leads to
hierarchies of lattice systems, of which the $r=1$ cases are the first
members.  So hierarcies also fit naturally into this framework.

\bn{\bf References.}
\medskip
\ni$^1$M.\ Toda, {\it Theory of Nonlinear Lattices}  (Springer, Berlin, 1988).
\sn$^2$M.\ J.\ Ablowitz and J.\ F.\ Ladik, J.\ Math.\ Phys.\ {\bf17},
        1011 (1976).
\sn$^3$L.\ D.\ Faddeev and L.\ A.\ Takhtajan, {\it Hamiltonian Methods in
        the Theory of Solitons} (Springer, Berlin, 1987).
\sn$^4$S.\ J.\ Orfanidis, Phys.\ Rev.\ D {\bf18}, 3822 (1978).
\sn$^5$M.\ Jimbo and T.\ Miwa,  Publ.\ RIMS {\bf19}, 943 (1983).
\sn$^6$R.\ S.\ Ward,  Phys.\ Lett.\ A {\bf199}, 45 (1995).
\sn$^7$R.\ Hirota,  J.\ Phys.\ Soc.\ Jpn.\ {\bf50}, 3785 (1981).
\sn$^8$H.\ W.\ Braden,  J.\ Phys.\ A {\bf30}, L485 (1997).
\sn$^9$N.\ Papanicolaou,  J.\ Phys.\ A {\bf20}, 3637 (1987).
\sn$^{10}$S.\ N.\ M.\ Ruijsenaars, Commun.\ Math.\ Phys.\ {\bf133}, 217 (1990).
\sn$^{11}$I.\ Merola, O.\ Ragnisco and G.\ Z.\ Tu, Inv.\ Prob.\ {\bf10},
         1315 (1994).
\sn$^{12}$B.\ Deconinck, Phys.\ Lett.\ A {\bf223}, 45 (1996).
\sn$^{13}$Yu.\ B.\ Suris, J.\ Phys.\ A {\bf30}, 1745 (1997).
\sn$^{14}$Yu.\ B.\ Suris,  A collection of integrable systems of the Toda type
        in continuous and discrete time, with $2\times2$ Lax representations.
        Preprint solv-int 9703004.
\sn$^{15}$Yu.\ B.\ Suris, Phys.\ Lett.\ A {\bf145}, 113 (1990).
\sn$^{16}$A.\ G.\ Izergin and V.\ E.\ Korepin, Lett.\ Math.\ Phys.\ {\bf5},
         199 (1981).

\bye